\documentclass[aps,pre,reprint]{revtex4-2}
\usepackage{mathtools}
\usepackage{amsmath}
\usepackage{amssymb}
\usepackage{graphicx}
\usepackage{dcolumn}
\usepackage{bm}
\usepackage{natbib}		
\usepackage{color}

\newcommand {\vp} {\varphi}

\def\w{\omega}

\newcommand{\av}[1]{\left\langle #1 \right\rangle}

\begin{document}
\title{Dynamics of oscillator populations with disorder in the coupling phase shifts}

\author{Arkady Pikovsky}
\affiliation{Institute of Physics and Astronomy, University of Potsdam,
Karl-Liebknecht-Str. 24/25, 14476 Potsdam-Golm, Germany}
\email{pikovsky@uni-potsdam.de}

\author{Franco Bagnoli}
\affiliation{Department of Physics and Astronomy and CSDC, University of Florence,  
via G. Sansone 1, 50019 Sesto Fiorentino (Italy)}
\affiliation{INFN, Section Florence}
\email{franco.bagnoli@unifi.it}

\date{\today}

\begin{abstract}
We study populations of oscillators,  all-to-all coupled by means of quenched disordered phase shifts. While there is no traditional synchronization transition with a nonvanishing Kuramoto order parameter, the system demonstrates a specific order as the coupling strength increases. This order is characterized by partial phase locking, which is put into evidence by the introduced correlation order parameter and via frequency entrainment. Simulations with phase oscillators,  Stuart-Landau oscillators, and chaotic Roessler oscillators demonstrate similar scaling of the correlation order parameter with the coupling and the system size and also similar behavior of the frequencies with maximal entrainment at some finite coupling.
\end{abstract}

\maketitle


\section{Introduction}
\label{sec:intro}
Synchronization phenomena in populations of oscillators have been widely studied since the pioneering works
of Winfree~\cite{Winfree-67} and Kuramoto~\cite{Kuramoto-75}. The Kuramoto model and its modifications, based on the phase dynamics, serve 
as paradigmatic models of synchronization (with a possibility, in many situations, of a fully analytical treatment). The relevance of the findings within phase dynamics is confirmed
by studies of realistic models of limit-cycle oscillators, where the amplitude dynamics is also included. If limit cycles are close to a Hopf bifurcation point, a universal normal form description can be adopted, which is expressed as an ensemble of Stuart-Landau oscillators, which also exhibits the synchronization transition~\cite{Hakim-Rappel-92,Nakagawa-Kuramoto-94}. Moreover, Kuramoto-type synchronization can be also observed in populations
of chaotic oscillators~\cite{Pikovsky-Rosenblum-Kurths-96}.

Effects of disorder on the synchronization properties of large populations of coupled oscillators 
are a subject of intensive current research. Randomness can be restricted to properties of individual units; e.g., already in the original formulation of the Kuramoto model~\cite{Kuramoto-75}, a distribution of natural frequencies was assumed. Also, the interactions between the oscillators can hold randomness, e.g., in the form of a random (weighted or unweighted) network~\cite{kalloniatis2010incoherence,chiba2018bifurcations,juhasz2019critical}.  In many cases, in the thermodynamic limit of an infinite number of units, a theory of synchronization transition can be constructed where the randomness is smeared out, and the averaged coupling, if attractive, 
leads to a synchronization transition~\cite{smirnov2023dynamics}. 

A special class of problems appears if the random coupling is maximally frustrating, i.e., if it disappears after a naive averaging. To the best of our knowledge, the first model of such type
has been suggested by Daido~\cite{Daido-92}, who considered a population of oscillators with mutual couplings
sampled from a Gaussian distribution with zero mean. Although such a model does not exhibit synchronization as a transition to a state with a non-vanishing order parameter, it demonstrates some entrainment properties.  By exploiting an analogy with spin glasses~\cite{fischer1993spin}, he found regimes with a slow relaxation to equilibrium. Furthermore, Daido found a volcano transition, where a distribution of random coupling inputs on oscillators changes its form from one with 
a maximum at zero to one with a maximum at a finite distance from zero. The Daido model and its modifications have been since intensively studied~\cite{Stiller-Radons-98,Daido-00,Stiller-Radons-00,Daido-18,Ottino-Strogatz-18,Pazo-Gallego-23,pruser2023nature}.

This paper focuses on another type of randomness in a population of oscillators, where couplings are characterized by random phase shifts~\cite{Park-Rhee-Choi-98}. If the distribution of the random shifts is nonuniform,
one obtains certain renormalized averaged interaction in the thermodynamic limit, which begets a standard synchronization transition~\cite{smirnov2023dynamics}. However, no standard synchronization transition occurs for a maximally frustrating disorder with a uniform distribution of the phase shifts in $[0,2\pi)$.
As we will show in this paper, a certain ordered state nevertheless appears, and the present work aims to characterize it.

The paper is organized as follows. In Section~\ref{sec:bm}, we will formulate the three basic models that will be explored numerically: phase oscillators, Stuart-Landau limit cycle oscillators, and Roessler chaotic oscillators. In Section~\ref{sec:char}, we introduce the tools by which the regimes will be characterized. In particular, we introduce a novel correlation order parameter and discuss its relation to the standard Kuramoto order parameter. Detailed numerical studies of populations of phase oscillators are presented in Section~\ref{sec:phosc}. Similar simulations for Stuart-Landau and Roessler oscillations reported in Section \ref{sec:slr} show, on the one hand, that the effects found for phase oscillators are reproduced for these more realistic models, while, on the other hand, novel features related to the amplitude dynamics appear. We conclude in Section \ref{sec:concl} with a comparison to Daido's disorder model. 

\section{Basic models}
\label{sec:bm}
In this section, we take three popular models of global coupling (Kuramoto-Sakaguchi 
model~\cite{Sakaguchi-Kuramoto-86}; system of Stuart-Landau oscillators~\cite{Hakim-Rappel-92};
system of chaotic Roessler oscillators~\cite{Pikovsky-Rosenblum-Kurths-96}), widely used for exploration
of the collective dynamics and of the synchronization transition, and extend them to the case of 
disorder in the phase shifts.
\subsection{Phase oscillators}
The standard Kuramoto-Sakaguchi model of $N$  oscillators with all-to-all couplings reads
\begin{equation}
\dot\varphi_k=\omega_k+f\sum_{j\neq k} \sin(\vp_j-\vp_k-\alpha),\quad k=1,\ldots,N\;.
\label{eq:ksm}
\end{equation}
Here $\varphi_k$ are the phases of the oscillating units, and $\omega_k$ are their natural frequencies.
Parameter $f$ is the coupling strength (for future compatibility, we do not normalize it by $N$).
Here, all mutual coupling terms are assumed to include the same phase shift $\alpha$. We introduce
a disorder in these phase shifts by assuming that phase shifts $\alpha_{jk}$ are independent, 
identically distributed random numbers (cf. Ref.~\cite{Park-Rhee-Choi-98}):
\begin{equation}
\dot\varphi_k=\omega_k+f\sum_{j\neq k} \sin(\vp_j-\vp_k-\alpha_{jk})\;.
\label{eq:rps}
\end{equation}
Furthermore, in this paper we assume the maximally possible disorder, where $\alpha_{jk}$ are uniformly
distributed in $[0,2\pi)$ (for non-uniform distributions, see theory in Ref.~\cite{smirnov2023dynamics}). Furthermore, we mostly concentrate on the asymmetric case $\alpha_{jk}\neq \alpha_{kj}$.

The main physical motivation behind the disordered phase shifts is that such shifts
can be associated with time delays in coupling. In particular, if the frequencies of all 
oscillators are close to a central frequency $\omega_0$, then the shifts can be 
expressed via the coupling time delays $\tau_{jk}$ 
as $\alpha_{jk}=\omega_0\tau_{jk}$, where $\omega_0$ is the central frequency. The basic assumption is that the time delays are smaller than the characteristic time of slow variations
of the phases; see Ref.~\cite{Izhikevich-98} for a precise formulation. Uniformity (or almost uniformity) of the distribution
of the phase shifts corresponds to a broad distribution of the time delays (a situation where delays in coupling 
obey a gamma distribution has been treated in Ref.~\cite{Lee-Ott-Antonsen-09}). We mention here that in addition 
to time delays in coupling, other factors may influence the phase shifts. For example, 
transmitting a signal from the driving to the driven unit may include filters with random characteristics.

We mention here that following the pioneering work of Daido~\cite{Daido-92}, one often considers an implementation of disorder
in the coupling, where the randomness is not in the phase shifts but in the coupling constants, 
see, e.g., Refs.~\cite{Stiller-Radons-98,Daido-00,Stiller-Radons-00,Daido-18,Ottino-Strogatz-18,Daido-18,Pazo-Gallego-23,pruser2023nature}. We discuss a possible relation to these models
in Section~~\ref{sec:concl}; here we only mention that the phenomenologies in the cases of the coupling strength disorder 
and the phase shift disorders are different. 

Below, we consider a Gaussian distribution of natural frequencies $\omega_k$. 
Since one can rescale time in Eq.~\eqref{eq:rps}, we will set the variance of this distribution to one.
Furthermore, to avoid potential finite-size effects due to realizations of the frequencies, we assign $\omega_k$ not
via random sampling, but via deterministic sampling using quantiles of the Gaussian distribution (see discussion
on the differences between random and deterministic samplings in Ref.~\cite{Peter-Pikovsky-18}).
Below, we will also consider a version of Eq.~\eqref{eq:rps} where all the oscillators have the same natural frequency 
(which is set to zero, $\omega_k=0$), but are subject to independent Gaussian white noise forces:
\begin{equation}
\begin{gathered}
\dot\varphi_k=\eta_k(t)+f\sum_{j\neq k} \sin(\vp_j-\vp_k-\alpha_{jk})\;,\\
\av{\eta_k}=0,\;\av{\eta_k(t)\eta_l(t')}=2\delta_{kl}\delta(t-t')\;.
\end{gathered}
\label{eq:rpsn}
\end{equation}
Here again, the intensity of noise is fixed via the time rescaling. 

In the case of deterministic identical
 oscillators, one can rescale the coupling parameter to one, and the problem is reduced to
\begin{equation}
\dot\varphi_k=\sum_{j\neq k} \sin(\vp_j-\vp_k-\alpha_{jk})\;,
\label{eq:rpsd}
\end{equation}
where the only parameter is the size of the ensemble $N$.

\subsection{Stuart-Landau oscillators}
A Stuart-Landau oscillator (SLO) is a prototypic model of a self-sustained oscillator, 
which includes both the phase and the amplitude dynamics. We consider in this paper a network
of $N$ SLOs with random phase shifts in coupling
\begin{equation}
\dot{a}_k=i\w_k a_k +\mu a_k(1-|a_k|^2)+f\sum_{j\neq k} e^{-i\alpha_{jk}} a_j\;.
\label{eq:cslo}
\end{equation}
Here $a_k$ are complex amplitudes of the units, and parameter $\mu$ defines the relaxation rate
of the amplitudes ($\mu^{-1}$ is the characteristic relaxation time of the amplitude dynamics).
In the limiting case of strongly stable amplitude, $\mu\to\infty$, the dynamics of the phases $\varphi_k=\text{arg}(a_k)$
follows Eq.~\eqref{eq:rps}; otherwise, the dynamics of the amplitudes influences that of the phases.

\subsection{Chaotic Roessler-type oscillators}
A chaotic Roessler oscillator~\cite{Rossler-76} (RO) possesses a well-defined phase variable; and an ensemble
of such systems with a Kuramoto-Sakaguchi-type coupling Eq.~\eqref{eq:ksm} demonstrates
a phase 
synchronization transition similar to the Kuramoto transition~\cite{Pikovsky-Rosenblum-Kurths-96}. Below we explore
the effect of randomness in the phase shifts in the coupling. For this purpose, we slightly modify the original 
Roessler system to obtain equations close to the dynamics of the SLO:
\begin{equation}
\begin{aligned}
\dot a&=\w(i a+d a-z)\;, \\
\dot z&=\w(b+z(\text{Re}(a)-c))\;.
\label{eq:ro}
\end{aligned}
\end{equation}
Here, $a$ is the complex ``amplitude'' of the oscillations (related to ``usual'' variables of the Roessler system as $a=x+iy$) , and $z$ is a variable,
the dynamics of which ensures chaotic ``saturation'' of the unstable oscillations of $a$. We introduce 
a parameter $\w$ responsible for the natural frequency as a common factor to avoid difficulties related to possible
bifurcations ``order-chaos'' in dependence on such a parameter (cf. a similar approach in Ref.~\cite{Rosa-Ott-Hess-98}).
The phase variable for the Roessler-type oscillator can be defined as  $\varphi_k=\text{arg}(a_k)$. The effective
dynamics of the phase is diffusive due to chaotic variations of the amplitudes.

We now introduce a global coupling in a population with different natural frequencies $\w_k$ like in \eqref{eq:cslo}
\begin{equation}
\begin{aligned}
\dot a_k&=\w_k(i a_k+d a_k-z_k) +f \sum_{j\neq k} e^{-i\alpha_{jk}} a_j\;, \\
\dot z_k&=\w_k(b+z_k(\text{Re}(a_k)-c))\;.
\end{aligned}
\label{eq:croes}
\end{equation}

\section{Characterizations of the collective dynamics}
\label{sec:char}

There are two ways to characterize synchronization in populations of oscillators: 
To follow the phase locking or the frequency entrainment. In many situations, these aspects have significant overlap,
but there are also cases where phase locking is accompanied by 
frequency anti-entrainment~\cite{Pimenova_etal-16,Goldobin_etal-17}. We will see that the two characterizations deliver complementary characterizations for the abovementioned ensembles.

\subsection{Correlation order parameter for phase locking}

Phase locking in a population of oscillators without disorder is typically characterized by
the complex Kuramoto order parameter (complex KOP)
\[
Z(t)=\frac{1}{N}\sum_{k=1}^N e^{i\varphi_k(t)}\;.
\]
At the synchronization transition, the absolute value of $Z$ starts to grow, and large
values of $|Z|$ indicate that the phases are concentrated in a small range (compared to $2\pi$).
As we will see below, this KOP is nearly vanishing (up to finite-size fluctuations) 
in the presence of disorder. Therefore, we suggest to characterize relations between the phases
with a complex phase correlation order parameter (complex COP) defined for the systems  Eqs.~\eqref{eq:rps},\eqref{eq:rpsn},\eqref{eq:rpsd} as
\begin{equation}
\begin{gathered}
Q(t)=\frac{1}{N}\sum_k Q_k(t)\;,\\
Q_k(t)=\frac{1}{N-1}\sum_{j\neq k} e^{i(\varphi_j-\varphi_k-\alpha_{jk})}\;.
\end{gathered}
\label{eq:copdef}
\end{equation}
We notice that the imaginary part of $Q_k$ is, up to a factor, the force acting on the oscillator $k$. 
The real part of $Q_k$ will be more important.
This quantity  characterizes the closeness (on average) of the phase $\varphi_k$ to the value
$\varphi_j+\alpha_{jk}$, which would be the locked state if only one coupling $j\to k$ would be present
(but this value is, of course, frustrated in the presence of all couplings).

Let us first demonstrate that complex COP and complex KOP are directly related in the case without disorder $\alpha_{jk}=0$,
i.e., for the standard Kuramoto setup: 
\begin{equation}
\begin{gathered}
|Z|^2=\frac{1}{N}\sum_{k=1}^N e^{-i\varphi_k(t)}\frac{1}{N}\sum_{j=1}^N e^{i\varphi_j(t)}=\\=
\frac{1}{N^2}\sum_{j,k} e^{i(\varphi_j-\varphi_k)}=
\frac{(N-1)Q+1}{N}\;.
\end{gathered}
\label{eq:kopcop}
\end{equation}
For large $N$, one thus has $|Z|^2\approx Q$ (note that in the ordered case, $Q$ is always real).

Below, in the applications of the complex COP to the disordered populations, we will perform additional averaging of $Q$ over time (for a particular
sample of phase shifts), and over realizations of disorder, and separate real and imaginary parts of the resulting complex quantity as
\begin{equation}
\av{Q}=C+iD\;.
\label{eq:realcop}
\end{equation}
We will refer to $C$ as the real correlation order parameter (real COP). Correspondingly, in a slight 
discrepancy with a usual definition of the (real) Kuarmoto order parameter, we will define it as $|Z|^2$, averaged over time and over the realizations
of disorder:  
\begin{equation}
R=\av{|Z|^2}\;.
\label{eq:realkop}
\end{equation}

The COP $Q$ can be directly applied to populations of SLOs \eqref{eq:cslo} and ROs \eqref{eq:croes}, 
provided the phases 
$\exp[i\varphi_k]= a_k/|a_k|$ are used as the observables. One can also define the corresponding correlation order parameters that 
include full amplitudes
\begin{equation}
\begin{gathered}
Q^{(a)}(t)=\frac{1}{N}\sum_k Q_k^{(a)}(t)\;,\\
Q_k^{(a)}(t)=\frac{1}{N-1}\sum_{j\neq k} a_j a_k^* e^{-i\alpha_{jk}}\;,
\end{gathered}
\label{eq:copadef}
\end{equation}
where $a_j$ are complex amplitudes either of SLOs or of ROs. Similar to \eqref{eq:realcop},
$\av{Q^{(a)}}=C^{(a)}+iD^{(a)}$.

\subsection{Frequency entrainment}

The definition of the average observed frequency for each oscillator in the population is straightforward (of course,
the average includes only time average, not averaging over the realizations of disorder):
\begin{equation}
\Omega_k=\av{\dot\varphi_k}\;.
\label{eq:frdef}
\end{equation}
This definition will be applied to all types of oscillators (phase oscillators, SLOs, and ROs).  
In the Kuramoto system without disorder, the spread of frequencies decreases beyond the synchronization transition. Moreover, in the
deterministic case, a portion (growing with the coupling strength) of oscillators is perfectly entrained so that 
their frequencies coincide. 

We notice here that the difference between the observed frequency $\Omega_k$ \eqref{eq:frdef}
and the natural frequency $\omega_k$ in \eqref{eq:rps} can be expressed as the average of the imaginary part of the local complex 
COP $(\Omega_k-\omega_k)\propto \av{\text{Im} (Q_k)}$. We also note that the entrainment of frequencies is a 
relevant characterization if the intrinsic frequencies $\omega_k$ are different; for the noisy identical oscillators
without the disorder, the frequencies are always equal.

\section{Phase oscillators}
\label{sec:phosc}

In this section, we apply the introduced above characterizations of synchrony in a population of oscillators to the phase
oscillators models Eqs.~\eqref{eq:rps},\eqref{eq:rpsn},\eqref{eq:rpsd}.
\subsection{Phase oscillators with a Gaussian distribution of frequencies}
\label{sec:dpo}
In exploring model \eqref{eq:rps} for different system sizes $N$, we select the frequencies
according to a Gaussian distribution with unit variance deterministically as corresponding quantiles. This is done
to avoid additional diversity due to the random sampling of frequencies. The only parameters of the model 
are the coupling strength $f$ and the system size $N$. The phase shifts in each run are chosen 
randomly according to a uniform distribution
$0\leq \alpha_{jk}<2\pi$, an averaging is performed over time and over different samples of phase shifts. 

\begin{figure}[!h]
\includegraphics[width=\columnwidth]{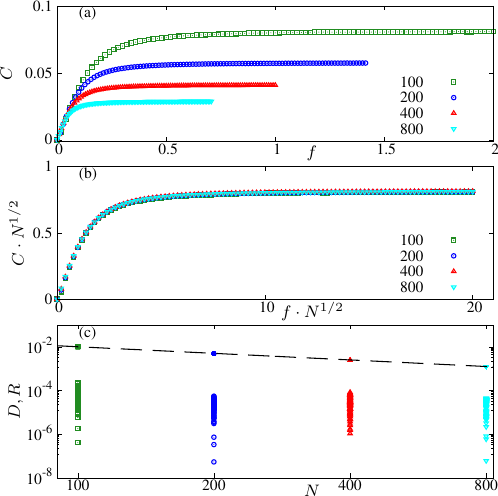}
\caption{Panel (a): real COP $C$ (see Eqs.~\eqref{eq:realcop},\eqref{eq:copdef}) for different system sizes in dependence on $f$. Panel (b): the same data as in panel (a), but in rescaled coordinates.
Panel (c): imaginary COP $D$ (see Eqs.~\eqref{eq:realcop},\eqref{eq:copdef}; scattered open markers in the bottom part of the panel) and $R$ (see Eq.~\eqref{eq:realkop}, filled  markers at the top part of the panel;
the values for different $f$ nearly overlap) for different $N$. The dashed line shows the law $\sim N^{-1}$ for the values of $R$.}
\label{fig:rps}
\end{figure}

Figure~\ref{fig:rps} presents the characterization in terms of the KOP and of the COP. The main finding is that while the value of the KOP $R$ 
defined in Eq.~\eqref{eq:realkop} exhibits finite size fluctuations $\sim N^{-1}$, which are practically independent on the coupling
strength $f$ (panel (c)), the value of the real COP $C$ defined in Eq.~\eqref{eq:realcop} exhibits a monotonic growth with a saturation at large $f$ (panel (a)).
At the same time, the imaginary  COP $D$ fluctuates close to zero (panel (c)).

In Fig.~\ref{fig:rps}(b) we also demonstrate, that the COP $C$ fulfills a scaling relation
\begin{equation}
C(f,N)=N^{-1/2}c(f N^{1/2})\;.
\label{eq:copsc}
\end{equation}
Remarkably, the COP demonstrates a continuous transition without any particular threshold value of the coupling. 

\begin{figure}[!h]
\includegraphics[width=\columnwidth]{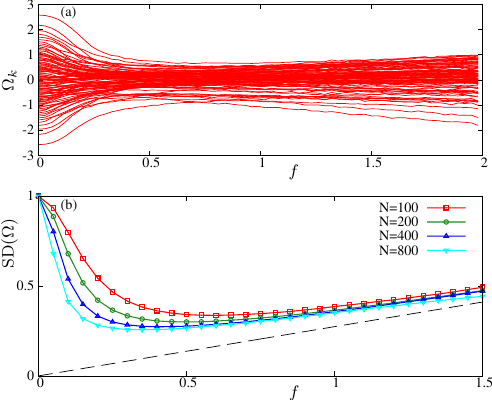}
\caption{Panel (a): Evolution of the frequencies $\Omega_k$ (see Eq.~\eqref{eq:frdef}) with coupling $f$ for one realization of disorder ($N=100$).
Panel (b): standard deviation of the frequencies (averaged over many realizations) vs $f$ for different 
$N$. The straight dashed line has a slope of $0.274$ (see discussion in text).}
\label{fig:rpsom}
\end{figure}

Next, in Fig.~\ref{fig:rpsom}, we explore the properties of the frequency entrainment. We show in panel (a) the dependencies
of frequencies $\Omega_k$ according to Eq.~\eqref{eq:frdef} for one realization of disorder in dependence on the coupling strength $f$,
and in panel (b) the standard deviations of the frequencies from the mean value, averaged for several realizations of disorder. One can see that
at small coupling strengths $f$, the dispersion of frequencies decreases (from the value $1$, defined according
to the distribution of natural frequencies), but after reaching a minimum, it starts to increase again, 
nearly linear in $f$. We will discuss this feature and explain the value of the slope of the dashed line in more detail in Section~\ref{sec:ipo} below.

\subsection{Noisy phase oscillators}
\label{sec:dpon}
\begin{figure}[!h]
\includegraphics[width=\columnwidth]{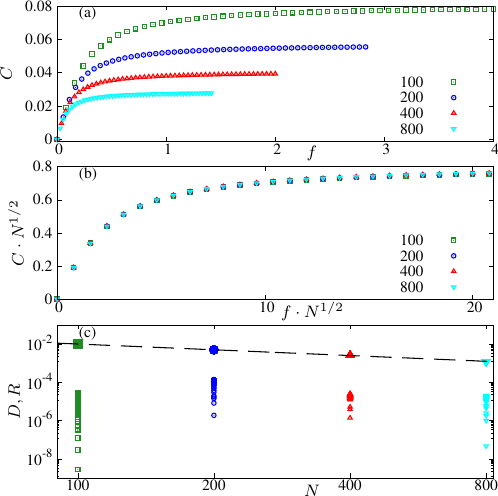}
\caption{The same as Fig.~\ref{fig:rps}, but for noisy oscillators Eq.~\eqref{eq:rpsn}.}
\label{fig:rpsn}
\end{figure}

\begin{figure}[!h]
\includegraphics[width=\columnwidth]{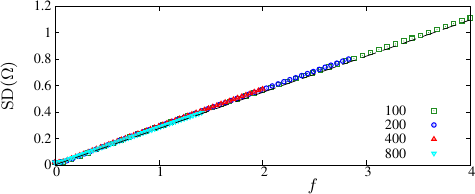}
\caption{Standard deviation of frequencies (averaged over many realizations) vs $f$, for noisy oscillators Eq.~\eqref{eq:rpsn}. The dashed line has a slope of $0.274$.
}
\label{fig:rpsnom}
\end{figure}

Here, we apply the same analysis as in Section~\ref{sec:dpo} above to noisy oscillators described by Eqs.~\eqref{eq:rpsn}.
First, we show in Fig.~\ref{fig:rpsn} the results for the order parameters characterizing phase coherence. One can see that 
there are no qualitative differences with the case of deterministic oscillators (Fig.~\ref{fig:rps}).  However, there are 
differences in the behavior of frequencies. Since the oscillators are identical,  the observed frequencies
$\Omega_k$ in the absence of coupling coincide. However, with the onset of the coupling, such differences do appear, and, as Fig.~\ref{fig:rpsnom} shows,
the standard deviation grows linearly with $f$. This amplification corresponds to the increase of the dispersion of the frequencies for
non-identical oscillators Eq.~\eqref{eq:rps} at large coupling strengths $f$ (Fig.~\ref{fig:rpsom}(b)).

\subsection{Identical phase oscillators}
\label{sec:ipo}

The results of Sections~\ref{sec:dpo},~\ref{sec:dpon} above show that at large coupling strengths $f$, a universal regime
appears, at which the real COP saturates at a value $C_{sat}\approx 0.8 N^{-1/2}$, and the dispersion of the 
frequencies grows proportional to $f$. In this state, one can neglect differences between the oscillators (either due to their
different natural frequencies or due to independent noise terms) and study a population of identical deterministic
oscillators as defined in Eq.~\eqref{eq:rpsd}. Because now the time is rescaled according to the coupling strength, a  constant dispersion
of the observed frequencies will explain linear growth of this dispersion with $f$ in setups where the coupling strength is not
rescaled, i.e., in Figs.~\ref{fig:rpsom}(b),\ref{fig:rpsnom}.

The only parameter of Eqs.~\eqref{eq:rpsd} is the population size $N$, what makes a statistical evaluation
of different characteristics easier. 

\begin{figure}[!h]
\includegraphics[width=\columnwidth]{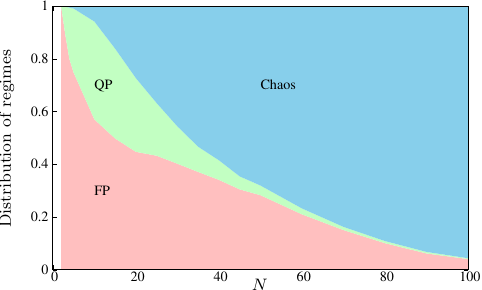}
\caption{Distribution of the regimes (FP: fixed point; QP: quasiperiodic) in dependence on $N$, 
for identical phase oscillators (statistics over  $10^4$ realizations of disorder).}
\label{fig:states}
\end{figure}

First, we would like to characterize the dynamical regimes. Computation of the maximal Lyapunov
exponent for the deterministic system \eqref{eq:rpsd} allows us to distinguish chaos and regular states. Because the system \eqref{eq:rpsd}
is invariant with respect to the shift of all phases, one Lyapunov exponent is always zero. To distinguish whether
a state with vanishing maximal Lyapunov exponent is a synchronous one where all the frequencies coincide or a quasiperiodic state 
where the frequencies are different, we calculated in each run the spread of the observed
frequencies \eqref{eq:frdef} as $\Delta\Omega=\Omega_{max}-\Omega_{min}$. Regimes where this spread nearly 
vanishes (we used a threshold $10^{-4}$
as a practical criterion) were identified as steady states (in some reference frame rotating 
with a common frequency of the oscillators). Otherwise, the states with nearly vanishing maximal Lyapunov exponent and 
$\Delta\Omega>10^{-4}$ were identified as quasiperiodic regimes. The probabilities of observing three possible regimes
depend on the ensemble size as depicted in Fig.~\ref{fig:states}. For $N=2$, for all phase shifts, the oscillators synchronize (as the theory of coupled oscillators predicts, because there is no frustration for two oscillators). At the same time, for larger $N$, quasiperiodic and chaotic regimes become possible. Finally, for large $N\gtrsim 100$, chaotic regimes dominate.

Next, we explored the statistics of the observed frequencies Eq.~\eqref{eq:frdef}  at large $N$, and found that
the distribution of these frequencies (when averaged over many realizations of disorder) is Gaussian with high accuracy.
Furthermore, the variance is practically the same in a large range of $N$. This observation explains the linear behavior
of the standard deviations of the frequencies in Figs.~\ref{fig:rpsom}(b) and \ref{fig:rpsnom}, and also the fact that 
these standard deviations
for different $N$ nearly coincide at a value $\approx 0.274$. This gives a prediction $\text{Std}(\Omega)\sim 0.274 f$ for large
$f$ in the presence of a distribution of natural frequencies or in the presence of noise; this prediction is shown with dashed lines
in Figs.~\ref{fig:rpsom}(b) and \ref{fig:rpsnom}.

Finally, we studied the statistics of the complex COP $Q_k$ defined in Eq.~\eqref{eq:copdef}. We have found 
that the values $\text{Re}(Q_k)$ and $\text{Im}(Q_k)$ are practically uncorrelated. The average of $\text{Im}(Q_k)$
is nearly zero, while the average of  $\text{Re}(Q_k)$ gives the averaged COP $C$, as discussed above. The variances
of  $\text{Re}(Q_k)$ and $\text{Im}(Q_k)$ are found to be $\sim N^{-1}$. Remarkably, the distribution
of  $\text{Re}(Q_k)$ is rather close to a Gaussian one, with small skewness values and excess kurtosis values.
The distribution of $\text{Im}(Q_k)$ is nearly symmetric with a small skewness, but the excess kurtosis values of $\approx 3$
indicate deviations from a Gaussian distribution.

\section{Stuart-Landau and Roessler-type oscillators}
\label{sec:slr}
This section aims to show that the essential effects described above for the phase oscillators
are also observed for SLOs and ROs. We do not repeat all the numerical setups but focus on 
ensembles with a distribution of natural frequencies without noise.

\subsection{Stuart-Landau oscillators}
Stuart-Landau oscillators have a parameter $\mu$ that governs the stability of the
limit cycle. For $\mu\to\infty$, a reduction to the phase dynamics Eq.~\eqref{eq:rps}
becomes exact. Thus, we focus mainly on additional effects appearing at small $\mu$.

\begin{figure}[!h]
\includegraphics[width=\columnwidth]{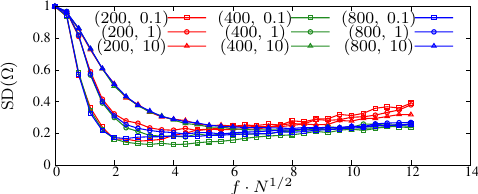}
\caption{Standard deviations of frequencies $\Omega_k$ vs $f\sqrt{N}$ 
for one realization of disorder and different $N,\mu$ (the values of these parameters are shown in brackets, e.g., $(200,0.1)$ denotes $N=200$, $\mu=0.1$) for SLO oscillators.}
\label{fig:slo_om}
\end{figure}

We start with presenting in Fig.~\ref{fig:slo_om} the evolution of the frequencies 
for an ensemble of SLOs 
\eqref{eq:cslo}, where natural frequencies are sampled according to a Gaussian distribution with
variance $1$. One can see that the evolution of standard deviation is very similar to that 
for phase oscillators (cf.~Fig.~\ref{fig:rpsom}): entrainment at small values of coupling and 
growth of the standard deviation for strong coupling. The dependence of these
effects on the parameter $\mu$ is only quantitative: for small values of $\mu$, the minimal standard deviation is achieved at smaller 
values of $f$ (see in Fig.~\ref{fig:slo_om} the curves with an open square marker, corresponding to $\mu=0.1$).

\begin{figure}[!h]
\includegraphics[width=\columnwidth]{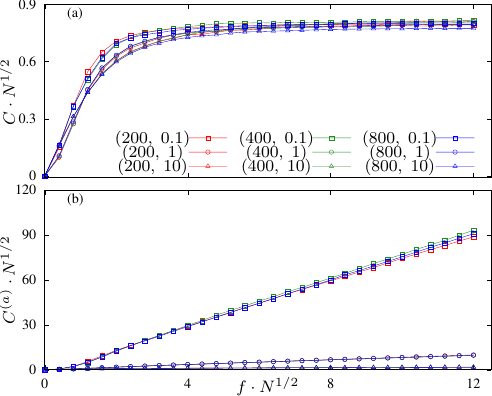}
\caption{Ensembles of SL oscillators. Panel (a): read COP $C$ according to Eq.~\eqref{eq:copdef} vs. $f$, in rescaled coordinates. 
Panel (b): real COP $C^{(a)}$ according to Eq.~\eqref{eq:copadef} vs $f$, in rescaled coordinates. The values of $N,\mu$ are marked like in Fig.~\ref{fig:slo_om}.}
\label{fig:SLop}
\end{figure}

As discussed above, for the COP, we now have two variants: one is based on the 
transformation to the phases and thus is mostly close to the COP for the phase oscillators 
(Eq.~\eqref{eq:copdef}); another one is based on the cross-correlations of the amplitudes 
(Eq.~\eqref{eq:copadef}). Both these COPs are depicted in Fig.~\ref{fig:SLop}, and they 
demonstrate very different
behaviors. While the phase-COP is practically $\mu$-independent, the amplitude-COP grows 
strongly with the coupling if $\mu$ is small. We explain this by noticing that the linear coupling
of the SLOs affects the stability of the equilibrium $a_k=0$ in a population of SLOs.
Indeed,  Eqs.~\eqref{eq:cslo} for small amplitudes $a_k$ can be written as a linear system
\begin{equation}
\dot{a}_k=i\w_k a_k +\mu a_k+f\sum_{j\neq k} e^{-i\alpha_{jk}} a_j\;,
\label{eq:cslol}
\end{equation}
and  stability is affected by the random coupling matrix $e^{-i\alpha_{jk}}$. According to the
circular law~\cite{Tao-Vu-10}, for large $N$, the eigenvalues of this matrix are
uniformly distributed in a disk with radius $\sqrt{N}$ around the origin. Thus, the largest
real part of the eigenvalues in Eq.~\eqref{eq:cslol} can be estimated as $\tilde\mu\approx \mu+f\sqrt{N}$. 
If we assume that this instability saturates at amplitude $|a|^2\approx \tilde\mu/\mu$, we conclude that
the average amplitude of oscillations, for large couplings, grows $\sim f\sqrt{N}/\mu$. This linear 
growth corresponds
to what is observed in Fig.~\ref{fig:SLop}, and explains why the amplitude-COP is not a proper quantity to characterize
phase coherence in the ensemble for small values of $\mu$.

\subsection{Roessler oscillators}
Here, we report on the effects of coupling with disorder in the phase shifts 
on the Roessler system \eqref{eq:croes}. The natural frequencies $\omega_k$ have been sampled from
Gaussian distribution with mean value $1$ (because the RO is not rotationally invariant, one cannot
set the mean frequency to zero by a transformation to a rotating reference frame, as it is 
possible for the phase oscillators and the SLOs) and standard deviation $0.1$. Other parameters were set to
$d=0.09$, $b=0.4$, $c=8.5$. We also note that because the disordered 
coupling enhances instability of the oscillations, for large enough values of $f$, no finite
regimes in system  \eqref{eq:croes} have been observed because a trajectory escaped to infinity. Thus, we report on the regimes 
for relatively small values of coupling strength $f$ only.

\begin{figure}[!h]
\includegraphics[width=\columnwidth]{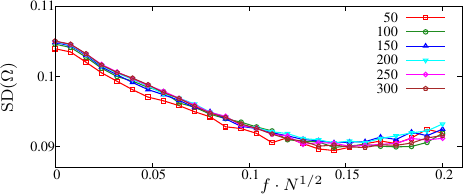}
\caption{Standard deviations of frequencies $\Omega_k$ vs $f\sqrt{N}$ for one realization of disorder and different $N$, for Roessler oscillators}
\label{fig:roes_om}
\end{figure}

The evolution of the standard deviations of the observed frequencies with $f$ is reported in
Fig.~\ref{fig:roes_om}. One can see that in the explored range of system sizes, this standard deviation
is a universal function of rescaled coupling $f\sqrt{N}$. Frequency entrainment is maximal at 
$f\sqrt{N}\approx 0.14$; however, the overall decrease of the standard deviation is less pronounced than for the phase oscillators 
and the SLOs (cf. Fig.~\ref{fig:rpsom}(b) and Fig.~\ref{fig:slo_om}).

\begin{figure}[!h]
\includegraphics[width=\columnwidth]{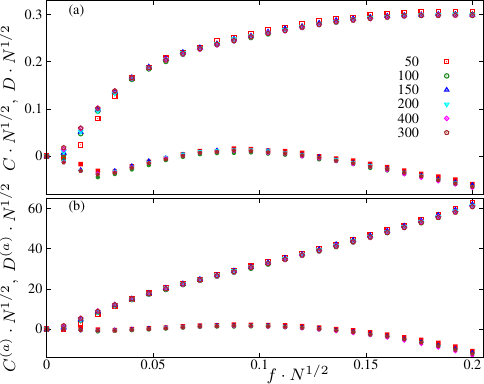}
\caption{Roessler oscillators. Panel (a):  real COP $C$ (open markers, upper set of points)
and  imaginary  COP $D$ (filled markers, bottom (closer to zero) set of points) according to
Eq.~\eqref{eq:copdef}, for different $N$. Panel (b): the same for the order parameter $Q^{(a)}$ (Eq.~\eqref{eq:copadef}). All coordinates are rescaled $\sim N^{1/2}$.
}
\label{fig:roes_cop}
\end{figure}

Calculations of the COP, depicted in Fig.~\ref{fig:roes_cop}, 
revealed the same scaling in dependence on the $f,N$ as in relation \eqref{eq:copsc}. The main 
difference between the corresponding results for the phase oscillators and the SLOs 
is that the imaginary part $D$ also deviates from zero. We attribute this to the fact that
the RO Eq.~\eqref{eq:croes}, due to the existence of the variable $z$, is non-invariant 
toward rotations of the phase $a\to a e^{i\theta}$. 

The difference in the amplitude of the values
of $Q$ and $Q^{(a)}$ has the same explanation as in the case of the SLOs above: disordered coupling contributes
to the instability of oscillations of the variable $a$, thus increasing the amplitude of the chaotic 
attractor, which contributes to the growth of $Q^{(a)}$.

\section{Conclusion}
\label{sec:concl}

We first summarize the results of our study. We have demonstrated that globally coupled oscillators
with a maximally frustrating disorder in the coupling phase shifts demonstrate a transition to a partially ordered
state. This state cannot be identified by the standard Kuramoto order parameter, 
which remains close to zero. We introduced a correlation order parameter, the calculation of 
which requires knowledge of the phase shifts, and demonstrated that it continuously 
grows with coupling
achieving a saturation level $\sim N^{-1/2}$ beyond a characteristic coupling strength $\sim N^{-1/2}$
(expression \eqref{eq:copsc}). 

The transition to order can also be characterized via frequency
entrainment; here, one does not need to know the phase shifts, and this characterization can be easily 
implemented in experimental setups. Coupling leads to a concentration of frequencies, which can be 
straightforwardly quantified by a reduced standard deviation of their distribution. 
Remarkably, there is an 
optimal coupling strength at which the concentration is maximal; beyond this strength, the 
frequencies start to spread again. The disorder-induced spread of
the frequencies in a population of identical oscillators explains the latter effect. 

We have demonstrated that the ordering behavior
described above holds for oscillators with a distribution of natural frequencies and noisy identical
oscillators. Furthermore, we showed the effect not only for the phase oscillators but
also for the Stuart-Landau oscillators and for the coupled chaotic Roessler systems. We focused on the case where the phase shifts are not symmetric (i.e., $\alpha_{jk}\neq \alpha_{kj}$). Selected simulations with
symmetric random phase shifts (i.e., $\alpha_{jk}= \alpha_{kj}$) demonstrated only minimal 
quantitative difference to the presented features.

We have explored populations of oscillators with sizes up to several hundred; the scaling
properties are well-established in this range. However, probably not all features can be downscaled
to small ensembles. The reason is that for small populations, regular quasiperiodic (and periodic for 
identical deterministic units) become dominant, as the calculations of the largest Lyapunov 
exponent show. However, for $N\gtrsim 100$, chaotic regimes dominate.

Next, we compare our findings with other results on the disordered populations of coupled oscillators
from the recent literature. We have found only paper~\cite{Park-Rhee-Choi-98} where a Kuramoto-type
model with random phase shifts has been explored. Adopting the replica method for noisy 
phase oscillators with a Gaussian distribution of natural frequencies, the authors
predicted for a maximally frustrating disordered case (phase shifts uniformly distributed in $[0,2\pi)$)
the appearance of a glass synchronization state if $fN^{1/2}$ exceeds a certain level, which 
nontrivially depends on the noise intensity. It is unclear if this ``glass synchronization'' state
corresponds to a saturated ordered regime observed in our study at large couplings 
(no numerics is presented in Ref.~\cite{Park-Rhee-Choi-98}); such a comparison 
deserves further investigation. 

In this respect, we also mention that we have not found any indication
of the extremely slow dynamics, typically associated with a ``glassy'' state. Such a slowing
has been observed in the Kuramoto model with another type of disorder, not in the phase shifts, but
in the coupling constants (which are sampled from a Gaussian distribution with zero mean), 
see Refs.~\cite{Daido-92,Stiller-Radons-98,Daido-00,Stiller-Radons-00,Daido-18,Ottino-Strogatz-18,Pazo-Gallego-23,pruser2023nature}. In the latter model also a sharp ``volcano transition'' has been observed; we have not found this transition in the random phase shifts model. This comparison shows that different types of disorder apparently lead to different types of the dynamics. Thus, it is appealing to look at combinational models, where both types of the disorder (in the phase shifts and the coupling constants) are present; this work is in progress.

\acknowledgments
We thank L. Smirnov, D. Pazo, I. Kiss, O. Omelchenko, and M. Rosenblum for valuable discussions. 
A. P. thanks the University of Florence for warm hospitality.

\bibliography{rps}

\end{document}